\documentclass[twocolumn,amsmath,showpacs,amssymb,aps,nofootinbib,floatfix]{revtex4-1}
\usepackage{epsfig,amsmath,amssymb}
\usepackage{dcolumn,color}
\usepackage{bm}
\usepackage{graphicx}

\newcommand{\blacktext}[1]{\color{black} #1 \color{black}}

\newcommand{\beq}{\begin{equation}}
\newcommand{\eeq}{\end{equation}}
\newcommand{\eqcomma}{\phantom{A},\phantom{A}}
\newcommand{\order}[1]{ \mathcal{O} \left( #1 \right) }
\newcommand{\ave}[1]{\left\langle #1 \right\rangle}

\begin{document}
\title{Causality and dissipation in relativistic polarizeable fluids}
\author{David Montenegro$^1$, Giorgio Torrieri$^1$}
\affiliation{$^1$ IFGW, Unicamp, Campinas, Brasil}
\begin{abstract}
Using field theory techniques we analyze the perfect fluid limit, defined as the fastest possible local equilibration, in a medium with polarizeability, defined as a non-zero local equilibrium partition of angular momentum into spin and vorticity.   We show that to restore causality a relaxation term linking vorticity and polarization, analogous to the Israel-Stewart term linking viscous forces and gradients,is required.    This term provides a minimum amount of dissipation a locally thermalized relativistic medium with polarizeability must have, independently of its underlying degrees of freedom.   For materials susceptible to spin alignment an infrared acausal mode remains, which we interpret as a Banks-Casher mode signaling spontaneous transition to a polarized phase.  With these ingredients, we propose a candidate for a fully causal Lagrangian of a relativistic polarizeable system near the perfect fluid limit, and close with some phenomenological considerations
\end{abstract}
\maketitle
\section{Introduction}
The dynamics of polarization density (spin alignment) in a fluid close to the ideal hydrodynamic limit, is currently not so well-understood.
Its development \cite{polhydro1,polhydro2,bec2,gauge,explanation,bec1,shu,vortsus,flork1,flork2,hongo} has been motivated by the experimental observation of transfer of angular momentum degrees of freedom from hydrodynamic vorticity to polarization \cite{nature} and long-standing thermodynamic phenomena such as the Barnett-effect \cite{barnett}.   
Beyond phenomenological issues, the fact that microscopic degrees of freedom have spin means their polarization susceptibility is generally non-zero even when the scale of microscopic thermalization is ``fast'', since spin carries a fraction of angular momentum also in thermal equilibrium, both local and global.

Including polarization into hydrodynamics could also help in having a well-defined stable theory.
Polarization could potentially regulate the fluctuation-driven instability that was pointed out in \cite{nicolis1,burch}:  Because of the fact that there is no mass gap for vortex formation, and there is also no vortex propagation, it is apparent, when hydrodynamics is looked at from a field theory perspective, that a hydrostatic ``vacuum'' is unstable against perturbations.   As argued in \cite{polhydro1}, such fluctuations, provided vortical susceptibility goes to zero as vorticity goes to zero, could give a soft mass gap to vortices fixing this instability.

A universal formulation of transport/hydrodynamics with polarization is still in development.

At one end, there are works with 1-particle Wigner functions (for instance see \cite{bec1,shu}) which can be used to model local equilibrium instantaneously, but most likely can not lead to a dynamical theory close to local equilibrium throughout its evolution, since vorticity-spin interactions require long-distance correlations of microscopic degrees of freedom \cite{polhydro1} which by definition factor out of one-particle Wigner functions.   

At the other end,a global equilibrium including spin-and vorticity is well-defined (for instance see \cite{bec2,vortsus}) but again making a transition to dynamics from such a global equilibrium ab initio state is non-trivial since the global equilibrium state with angular momentum is non-extensive, it can not be split up into many local equilibrium ``cells''.

A version of relativistic hydrodynamics which incorporates local equilibrium and microscopic polarization is conceptually challenging \cite{polhydro1,polhydro2,flork1,flork2}, since quite a lot of characteristics we usually associate with the ideal fluid limit (vorticity conservation, isotropy, coarse-graining) applies in a very different way when collective angular momentum excitations can be transferred to microscopic spin degrees of freedom.   As shown in \cite{polhydro1,polhydro2}, as dynamics is not dictated by conservation laws when polarization is present, lagrangian techniques \cite{nicolis1,nicolis3,gthydro,ushydro,gripaios} provide a way to deal with these issues.

However, as shown in \cite{polhydro2} a problem with causality remains: the mixing the sound and the vortex mode driven by polarization induces a generally acausal dispersion relation.   This latter issue is a manifestation of Ostrogradski's theorem \cite{limostro,ushydro} since a lagrangian with vorticity-polarization coupling depends on acceleration.

What this shows is that \cite{ushydro}
\begin{itemize}
\item New variables are necessary to avoid Ostrogradski's instabilities
\item These new variables must have a purely dissipative dynamics, to recover global equilibrium after a small perturbation.  This can be done within the lagrangian formalism using the doubled variables techniques \cite{grozdanov,galley,ushydro}
  \end{itemize}
In the rest of this work we shall therefore identify these new variables (basically, vorticity and polarization are decoupled in the transient dynamics), assume dissipative corrections, and show that indeed the problem with causality is resolved.

The relaxation time thereby obtained, however, unlike in the Israel-Stewart case, corrects a previously non-dissipative system and hence introduces a minimal amount of dissipation, dictated solely by causality and a non-zero polarizeability.  
In other words causality requires a bottom-up lower limit of thermalization of a relativistic fluid whose microscopic constituents have spin.  
\section{Polarization and dissipation in a causal field theory of hydrodynamics \label{viscfield}}
\subsection{The Lagrangian and Equations of motion}
Let us briefly recap the approach of \cite{polhydro1,polhydro2}, where the formalism we use is
introduced, explained and justified in detail:  
There, following the formalism of \cite{nicolis1} we
construct a lagrangian which contains the information of the equation of state, including an entropy term derived from the fluid Lagrangian coordinate degrees of freedom $\phi_I$
\begin{equation}
b = \left( \det_{IJ} \left[ \partial_\mu \phi_I \partial^\mu \phi_J \right] \right)^{1/2}
  \end{equation}
as well as the polarization tensor $y_{\mu \nu}$, a chemical potential analogue which however transforms as a vector in the co-moving frame \cite{polhydro1}.
$y_{\mu \nu}$ is constructed to be a generalization of the chemical potentials in \cite{nicolis3} from an internal $U(1)$ symmetry to a spacetime $SO(3)$ one,
\blacktext{
  \begin{equation}
    \label{generat}
  y_{\mu \nu} \sim  u_\beta \partial^\beta \left( \sum_i \theta_i(\phi_I) \hat{T}_i  \right)
  \end{equation}
  where $\theta_i$ are local phases, depending on the coordinates, and $\hat{T}_i$ are the generators}.   Since spin density is not conserved, $y_{\mu \nu}$ is an auxiliary field (rather than $\phi_I$ becoming a matrix in $y-$ space, as in \cite{nicolis3}), which forms a global $SO(3)$-invariant lagrangian together with the $\phi_I$ fields.  $\phi_I$ locally preserves $SO(3)$, $y$ does not.

Note that it represents an ``average of many particles'' per volume cell.  Hence, commutation relations, representation, and phases (the latter distinguish $SU(2)$ from $SO(3)$) are irrelevant.  The different ``phases'' represent the ``chemical potential'' of that component of the spin orientation in the co-moving frame. Weather the spins aligned are of fermions or bosons, and their respective spin, modifies the equation of state but 
is irrelevant to the symmetries of $y_{\mu \nu}$ as long as the number of particles in each volume cell $\gg 1$.  

$y_{\mu \nu}$ breaks local isotropy explicitly, but does not vanish at thermodynamic equilibrium, and hence should be present in the ''ideal fluid limit'' where the timescale of local equilibration vanishes.   The breaking of isotropy is similar to that in magnetic materials, be they ferro-magnetic (where it is spontaneous, as the minimum energy state is aligned in spin) or antiferromagnetic (where it generally occurs at angular momentum densities high enough to align spin, otherwise the lowest energy configuration has no magnetization).   However, this is not magnetism, since it is driven by the presence of angular momentum in the fluid rather than magnetic interactions, and will have the same sign for particles and anti-particles (as is experimentally shown to happen in heavy ion collisions).      Such a substitution of angular momentum for magnetic fields has long been known as ``Barnett effect'' \cite{barnett}.   
Indeed, in the case where vortical susceptibility is calculated explicitly \cite{vortsus}, the expression for magnetic and vortaic susceptibility parallel each other, suggesting the dynamics is the same up to $C$ symmetry.
In a fluid with no chemical potential one expects the spin alignment will not
produce a magnetic field (since the magnetic moment of particles and antiparticles is opposite), but it will break isotropy and take angular momentum out of vorticity and vice-versa.

For small polarizations, the
lagrangian will reduce to the form
\begin{equation}
  \label{lagdef}
 {\cal L} = F(b,y) = F(b(1-c\, y_{\mu\nu}y^{\mu\nu})), 
\end{equation}
 $c>0$ implies the material is analogue to a ferromagnet, with the potential to get spontaneously polarized (spin alignment lowers the free energy).  $c<0$ is equivalent to an antiferromagnet, with the ground state resisting spin alignment (polarization increases the free energy all other parameters being equal).  

  Both cases are realized in nature ($c$ is related to the susceptibility calculated in \cite{vortsus}), and could correspond to systems with ideal-fluid behavior.     We shall call these analogues to ferromagnets and antiferromagnets driven by vorticity ``ferrovortetic'' and ``antiferrovortetic'' materials \footnote{To distinguish the two effects, one needs mobile charge carries of both signs, which does not generally happen in non-relativistic magnets.   If the forces aligning spins are exchange forces rather than magnetic ones, it might be that what we know as some ferromagnets are actually ``ferrovortets''}..

Note that since $c$ can depend on temperature, ``antiferrovortetic'' here could just as well mean a vortetic material whose microscopic degrees of freedom are above the equivalent of the ``curie'' temperature, since no order parameter is implied\footnote{An antiferromagnets's usual order parameter is susceptibility, which jumps from zero in an antiferrovortetic phase to non-zero in a paramagnetic phase.  But we are not considering magnetohydrodynamics here, magnetic fields are assumed to be zero, so the order parameter is irrelevant}.  It just means that at local microscopic equilibrium spin angular momentum density vanishes.

For a well-defined local equilibrium, i.e. the absence of non-hydrodynamic microscopic ``spin-wave'' modes, we need vorticity and polarization to be parallel  \cite{polhydro1}, in other words
\begin{equation}
  \label{parall}
  y_{\mu\nu} = \chi(b,\omega_{\mu\nu}\omega^{\mu\nu})\omega_{\mu\nu}, 
\end{equation}
where the relativistic vorticity \cite{rezzolla} includes the enthalpy $w$ 
\begin{equation}
\label{vortdef}
\omega_{\mu \nu} = 2 \nabla_{[\mu} w u_{\nu]} 
\end{equation}
\[\ = 2 w \left( \nabla_{[\mu}u_{\nu]} - \dot{u}_{[\mu} u_{\nu]} +  u_{[\mu} \nabla_{\nu]} \ln w \right) \simeq  \beta  \nabla_{[\mu}u_{\nu]} \]
Here it is important to note that the distinction between ``thermal'' (Eq. \ref{vortdef}, whose circulation is conserved in relativistic ideal fluids) and ``kinematic'' circulation ($\nabla \times u_\mu$, whose circulation is conserved in incompressible non-relativistic ideal fluids) \cite{rezzolla,bec1}
can, to linear order in the equation of motion, be reabsorbed into the definition of $\chi$ (Eq. \ref{parall}) in terms of the Equation of state, since around a hydrostatic background the perturbation in enthalphy $w$ follows the perturbation in velocity
\begin{equation}
\label{kinematic}
\partial_\mu \delta \left( wu_\nu \right) \sim F(T) \partial_\mu \delta u_\nu + \order{(\delta u)^2}
\end{equation}  
where $F(T)$ is a function of temperature.

Since Kubo formulae are defined in terms of linear dispersion relations\footnote{This is a corollary of the assumption the system is close to local equilibrium. Note that equilibrium here must be local, since global equilibrium with angular momentum leads to rotation \cite{bec2}}, and since the definition of $\chi$ is arbitrary in this work, we can ignore the distinction between kinematic and thermal vorticity.

Then, provided polarization and vorticity are parallel \cite{polhydro1} (we shall refer to $\chi$ as the proportionality constant) the Lagrangian becomes a Legendre transform of the Energy density, analogously to the case with chemical potential \cite{nicolis3}.  The linearized dispersion relation derived from this lagrangian is however generally acausal \cite{polhydro2}.

As shown in \cite{polhydro2}, constructing equations of motion out of the Lagrangian in Eq. \ref{lagdef} and expanding around the hydrostatic limit leads to causality violation.
Proceeding from the conclusion of \cite{polhydro2} and the insight of \cite{ushydro,israel}, we consider Eq.\ref{parall} to be an asymptotic limit of a relaxation Maxwell-Cattaneo type equation \cite{israel}, 
\begin{equation}\label{IS}
\tau_Y \partial_\tau \delta Y_{\mu \nu} + \delta Y_{\mu \nu} = y_{\mu \nu} =  \chi(b,w^2) \omega_{\mu \nu}   
\end{equation}
Where non-equilibrium polarization, represented by a ``Magnon'' tensor $Y_{\mu \nu}$ \blacktext{with the same symmetry properties as those of $y_{\mu \nu}$ (Eq. \ref{generat})} ), evolves to its equilibrium value in a dissipative manner from arbitrary initial conditions.  $\tau_Y$ is related, via an analogous equation to the Kubo formula \cite{coming} (NB \cite{gauge}), the two-point function between polarization and vorticity, in time as vortices do not propagate
\begin{equation}
  \label{tauydef}
  \tau_Y \propto \lim_{w\rightarrow 0} \frac{1}{w} \mathrm{Im} \int d^3x \theta(t) \ave{ \left[ y_{ij}(t,\vec{x}), y_{ij}(0) \right]} \exp\left(  iwt \right)
\end{equation}
The real part will be, as usual, proportional to $\chi(b,\omega_{\mu\nu}\omega^{\mu\nu})/\beta$ while the imaginary part will be dissipative.  Both,as transport
coefficients, are functions of temperature, vorticity and perhaps chemical potentials.   
Formulae of this type should arise from a generalization of the identities derived in \cite{boulware} for theories admitting a breaking of isotropy due to polarization.  This will be done explicitly in a forthcoming work \cite{coming}.

Note that this is a first rather than a second order gradient term, unlike in the case of the Israel-Stewart relaxation time \cite{mooretaupi}, as is expected since here, unlike in the Israel-Stewart case, the limit of relaxation is ideal rather than dissipative.

This equation can be easily obtained from the Lagrangian formalism \cite{ushydro} \blacktext{ via the doubled variables technique \cite{galley,grozdanov}, where two copies of the theory are present (variable $X$ becomes $X_\pm$ in the notation of \cite{ushydro}) and, once a direction of time is chosen, dissipative terms are represented by terms $\sim X_+ X_-$ .   In \cite{ushydro} we have formulated a Lagrangian describing a Maxwell-Cattaneo equation, with an asymptotic relaxation of the dissipative part of the energy momentum tensor to viscous forces.

Hence we define non-equilibrium polarization degrees of freedom $Y_{\mu \nu}$, having the same symmetries as Eq.\ref{generat},and  a lagrangian of the same form as the dynamics of $\Pi$ in Eq. 36 of \cite{ushydro} 
\begin{equation}
  \label{genlag}
L = F(b(1-c\, y_{\mu\nu}y^{\mu\nu})) + \mathcal{L}_{IS - vortex}
  \end{equation}
\[\ \mathcal{L}_{IS - vortex} = \frac{1}{2} \tau_Y ( \, Y^{\mu\nu}_{-} \, u^{\alpha}_{+} \partial_\alpha Y_{\mu\nu +} - Y^{\mu\nu}_{+} u^{\alpha}_{-} \partial_\alpha Y_{\mu\nu -} ) + 
\]
\begin{equation}
+ \frac{1}{2} Y_{\mu\nu \pm}  Y^{\mu\nu}_{\pm}+ Y^{\mu\nu}_{\pm}\left( \chi(b,w^2) \omega_{\mu\nu} \right)
\end{equation}
such a Lagrangian, just like the Israel-Stewart Lagrangian, is free of Ostrogradski instabilities.   We shall proceed from the equation of motion (defined either in terms of any combination between $Y_+$ and $Y_-$ as per the Closed Time Path (CTP) symmetry illustrated in \cite{ushydro,grozdanov}.)}

One could worry about the universality of this choice, as opposed, for example, of writing a general Lagrangian in terms of magnon/spinwave degrees of freedom.
Magnons after all generically appear as free massless particles in all materials with spontaneously broken isotropy.  In a generic theory incorporating fluid dynamics with vorticity and magnon kinetics, the distribution of angular momentum between vorticity and isotropy in each cell will not follow local equilibrium.

As our theory is built around the local equilibrium assumption,  Eqs. \ref{IS} and \ref{genlag} give magnons a purely dissipative dynamics coupled only to collective degrees of freedom with angular momentum.
This is equivalent to assuming the effect of magnon-magnon interactions is so strong  as to ''quickly'' reach the state of local maximum entropy \cite{polhydro1}.
 The alternative (for example, adding a non-dissipative kinetic term for $Y_{\mu \nu}$ in the Lagrangian) would necessitate calculating transport properties for magnons from this Lagrangian.   i.e. Entropy is not guaranteed to be at a local maximum after dissipation, and the resulting lagrangian would become a microscopic Lagrangian to be coarse-grained.  If Eq. \ref{genlag} leads to causal dynamics then, close to the ideal fluid limit, this is what it will coarse-grain to since additional terms would contain more derivatives and a lack of local entropy maximization.   Causality is what we aim to test for in this work.  

We note that we linearize around the hydrostatic limit, under the physically reasonable assumption that any perturbation will be linear when one looks sufficiently in the beginning.  However, the stability of equations of the type of Eq. \ref{IS} has been established in a wider context \cite{jorgeis}.

In this regard, we note that an ``inverted'' relaxation equation 
\[\
\tau_Y \partial_\tau \delta \omega_{\mu \nu} + \delta \omega_{\mu \nu} =  \chi(b,w^2)^{-1} y_{\mu \nu}   
\]
would, according to the reasoning in \cite{polhydro1}, be necessary to resolve the vortical instability noted in \cite{nicolis1,burch}.  However, non-equilibrium vorticity is ill-defined without viscous Israel-Stewart terms, hence we do not see a coherent way to define such an inverted equation in the ideal hydrodynamic limit.   We shall therefore proceed with Eq. \ref{IS}, valid since in the linear regime fluctuation-dissipation guarantees the two approaches are equivalent.
\subsection{Causality analysis of perturbations around a hydrostatic limit}
Considering  a system without further parameters, i.e. without chemical potential, shear and bulk viscosity will give us dissipative modes in $Y^{\mu \nu}$ and sound and vortex modes due to EoS.
 Following the prescription of \cite{nicolis3}, the field $\phi^I$ describe a fluid out-of-equilibrium an general expansion can be made from the hydrostatic coordinates $\phi^I=x^I$
\begin{equation}
  \phi^I(x) = x^I + \pi^I + \frac{1}{2!} \pi \cdot \partial \pi^I + \frac{1}{3!}\pi \cdot \partial(\pi \cdot \partial \pi) + \order{\pi^4}
\end{equation}
where $\pi_I$ carry linearized sound/vortex perturbations.  Note that, as mentioned previously (Eq. \ref{kinematic}) changing the definition of vorticity from kinematic to thermal will, when the lagrangian is expanded in $\pi_I$, change the coefficients of order $\order{\pi^2}$ by terms depending on temperature only.
Only $\order{\pi^3}$ terms (the self-energies and ''three-point functions'') will directly feel the difference between the two definitions.  This is not relevant for the conclusions of this paper, although will affect how the system responds to thermal fluctuations (see discussion around Eq \ref{effz} in the conclusion), since the self-energies of the two vortices renormalizing $\chi$ will generally be different for the thermal and kinematic case.

The equation of motion to a general polarization from Euler Lagrange equation becomes
\begin{eqnarray}\label{eom}
2 c \partial_\mu \partial_\nu \bigg(  Y^{\rho \sigma}  \frac{\partial Y_{\rho \sigma}} {{\partial(\partial_\mu\partial_\nu \pi^I)}}\bigg)  =  A \big(  c^2_s \partial_I [\partial \pi]-\ddot \pi^I  \big)
\end{eqnarray}
with $ c^2_s = \frac{F^\prime(b_o) + 2 b_o F^{\prime\prime}(b_o)}{ F^\prime(b_o)}\eqcomma A= b_o F^\prime(b_o)  $ and $\partial \pi=\partial_I \pi_J, \ [\partial \pi] = \partial_I \pi^I $ (using the notation in \cite{gripaios}).  

To linear order Fluctuations of field could be written as
\begin{equation}
  \label{soundmodes}
  \vec{\pi} = \vec{\pi}_T + \vec{\pi}_L = \vec{\nabla} \Phi^I (\vec{x,t}) +  \vec{\nabla} \times \vec{\Omega}(x,t)
\end{equation}
where $\pi_L$ usually parameterize a sound wave, a deformation of coordinates $\phi_I$ parallel to the perturbation while $\pi_T$ is a vortex, in the direction perpendicular  to propagation of sound.    Because of sound-vortex mixing, $k\ne 0$ for $\pi_T$.
Polarization terms $Y^{\mu \nu}$, once relaxation terms Eq. \ref{IS} are included, will propagate differently from sound and vortices.    Thus, the sound potentials in Eq. \ref{soundmodes} can be Fourier-expanded separately
\vspace{3mm}
\begin{equation}
\left(
\begin{array}{c}
\Phi \\
\Omega\\
\end{array}
\right) =   \left(
\begin{array}{c}
\Phi_0 \\
\Omega_0\\
\tilde{Y}_0
\end{array}
\right) \exp \left[  i \left( w_{L,T,Y} t -\vec{k}.\vec{x}  \right)\right]
\end{equation}
We can now use a trick analogous to that used in \cite{romdev} to invert equation \ref{IS}. 
The Left Hand Side of Eq. \ref{IS} becomes, in Fourier space ($\eta_{I\mu}$ are the metric components),

\[\
  \chi(b_o,0) +  \left\{  - b_o \chi^\prime(b_o,0) i \, k^I \, ( \pi_L ) +  ( \omega_L^2  - c^2_s  (k_L^I)^2 )  (\pi_L)^2 \right\} + \nonumber
  \]
  \[\ \left\{ b_o \chi^\prime(b_o,0)  \frac{1}{2}[k^J (\pi^I_T) k^I (\pi^J_T)]  \right\}- \omega^2 \chi(b_o,0) \times \]
\[\
 \times    \bigg\{   \delta_\mu^P  k_{(P\Omega)} \, \delta_\mu^Q  k_{(Q\Omega)} \, ({\bf k}\times {\boldsymbol \Omega^0})^I \, ({\bf k}\times {\boldsymbol \Omega^0})^I +  \]
\begin{equation}
 k^I_\Omega \, k^J_\Omega \, ({\bf k}\times {\boldsymbol \Omega^0})^J \, ({\bf k}\times {\boldsymbol \Omega^0})^I  \bigg\} 
\end{equation}
where the first two terms $\sim \pi_L^2$ represent, respectively, the diffusion (imaginary) and real (sound mode), the term $\sim \pi_T^I \pi_T^J$ is the vortex (transverse excitation) and the last term represents the Israel-Stewart mode relaxing to a vortex.
Note that sound waves have the speed \[\ c^2_s = \frac{b^2_o \chi^{\prime\prime} (b_o,0) + \frac{b_o}{2} \chi^\prime(b_o,0)}{\frac{b_o}{2} \chi^\prime(b_o,0)}\] and the imaginary part has non-propagating mode. Now, we will see the first order and second order expansion of $\omega^{\mu\nu}$
\begin{equation}
\frac{\partial \tilde{Y}^{\mu\nu} }{\partial(k_\alpha k_\beta \pi_T^L)} = 
\frac{2}{(1 + i \omega_Y\tau_Y)}  \chi(b_o,0) \bigg\{ \eta_{P \mu} \eta_{\nu Q} \delta^0_\alpha \delta^Q_\beta \delta^P_L  \bigg\}
\end{equation}
note that, as conjectured in \cite{polhydro2}, equation \ref{IS} now only has gradients up to order two, in contrast to the equations of motion of a fluid where polarization and vorticity align automatically. Ostrogradski's instabilities therefore should be absent.   

To this order,$\tilde{Y}$ is completely determined, just like the non-equilibrium part of the stress-energy tensor in \cite{romdev}.
The full dispersion relation for $\tilde{Y}$ is
\begin{widetext}
\begin{eqnarray}\label{pol}
\tilde{Y}^{\mu\nu} &=& \frac{1}{1 + i \omega_Y \tau_Y}  \bigg\{ \underbrace{\chi (b_o,0) g_{I[\mu} g_{\nu]J} \left[ \omega k^J_\Omega ({\bf k}\times {\boldsymbol \Omega^0})^I \right] e^{i\omega_{T} t -i{\bf k_T}\cdot {\bf x}}}_{ \sim \ \pi_T} \\
&& \underbrace{- i \, b_o \chi^\prime (b_o,0)  \, k^I \, ( k^I \Phi^0)  \bigg\{ g_{I[\mu}g_{\nu]J} \left[ \omega k^J_\Omega ({\bf k}\times {\boldsymbol \Omega^0})^I \right] \bigg\} e^{i(\omega_{L} + \omega_{T}) t - i( {\bf k_T}\cdot {\bf x} + {\bf k_{L}}\cdot {\bf x}) } }_{\sim \ \pi_L \pi_T}    \nonumber \\
&& \quad \underbrace{- i \chi(b_o,0)\bigg\{ g_{I [\mu} g_{\nu] J} (\omega) (k^J_\Omega k_{P\Omega} ) ({\bf k}\times {\boldsymbol \Omega^0})^P ({\bf k}\times {\boldsymbol \Omega^0})^I + g_{I[\mu} g_{\nu] J}  (\omega \, k^J_\Omega \, k^P_{\Omega}) ({\bf k}\times {\boldsymbol \Omega^0})^I ({\bf k}\times {\boldsymbol \Omega^0})^J  \bigg\}   e^{i\omega_{T} t - i{\bf k_T}\cdot {\bf x}}}_{\sim \ \pi_T \pi_T }  \nonumber
\end{eqnarray}  
\end{widetext}  
Therefore, only the first term of equation above is first order at $\pi$. The second term represent an interaction between sound waves and vorticity, while the final term is second order at vorticity perturbation.

\begin{figure}[h]
\includegraphics[height=0.25\textheight]{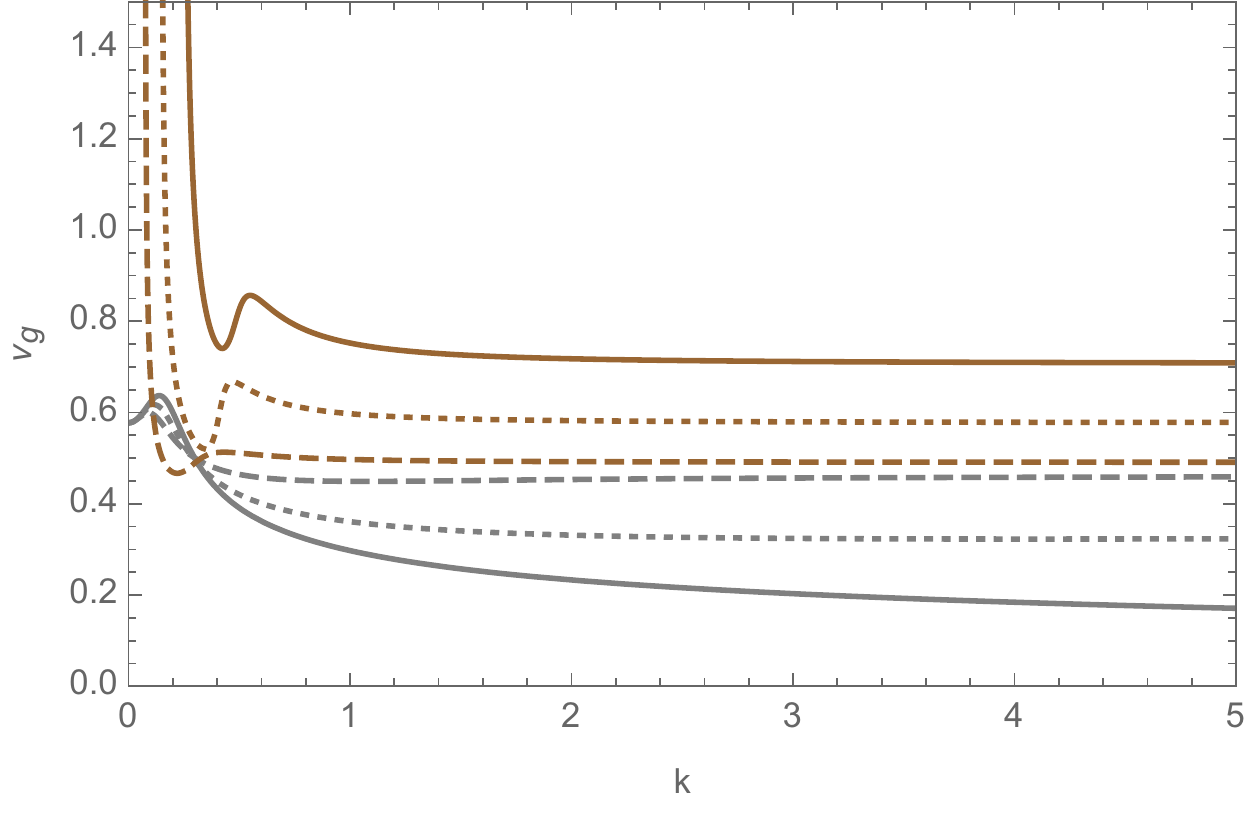}
\includegraphics[height=0.25\textheight]{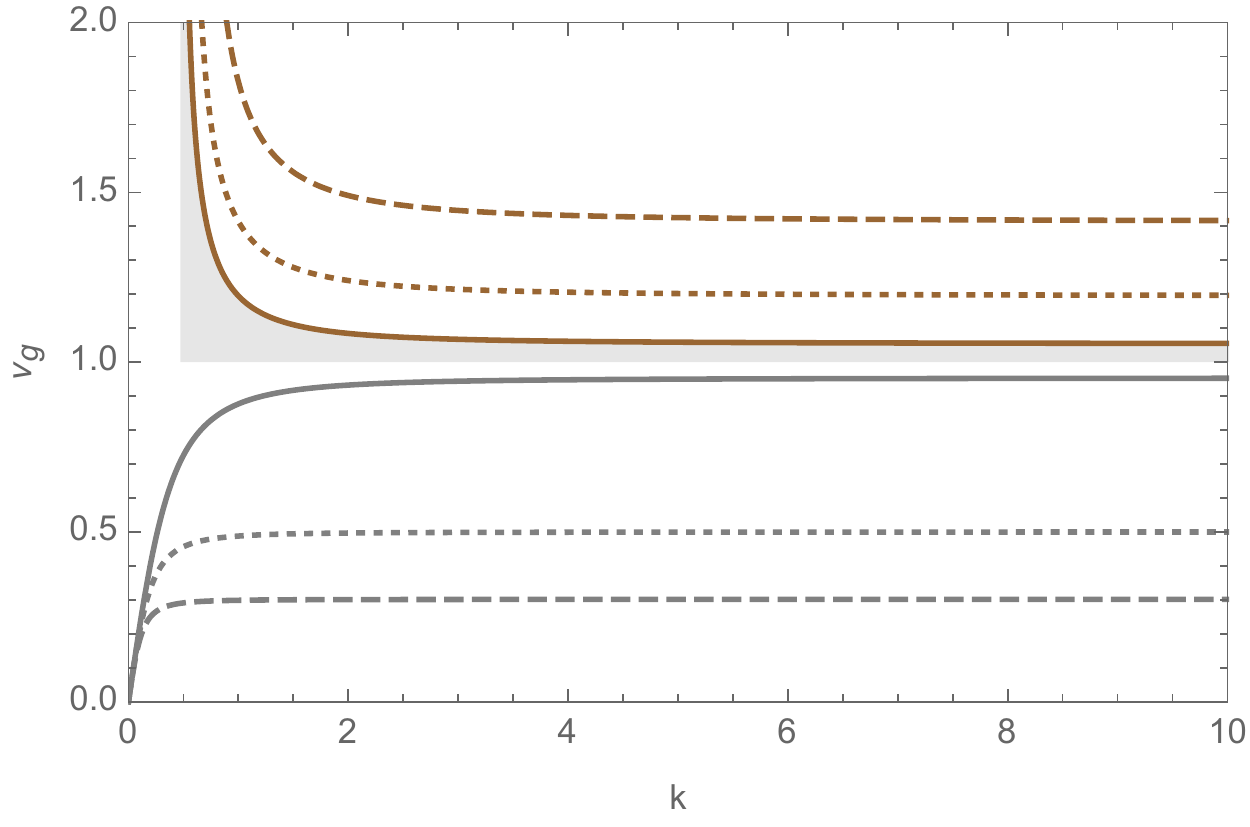}
\caption{\label{dispfig} The longitudinal and transverse dispersion relations for various entropy densities.
Non-causal region is shaded. The grey lines correspond to $c<0$ and Brown lines is $c>0$. The top figure shows sound modes. The full, dotted, and dashed lines are $\frac{\tau^2_Y }{(B/A)} = \left\{ 3, 5, 10 \right\}$, respectively, for both gray and borrow color. The bottom figure is transverse mode where gray: full, dotted, and dashed line are $\frac{\tau^2_Y }{(B/A)} = \left\{ 0.1, 0.3, 0.5\right\}$, respectively. Borrow: full, dotted, and dashed line are $\frac{\tau^2_Y }{(B/A)} = \left\{0.1,3,10 \right\}$ , respectively.   }
\end{figure}

Plugging the expression for $\tilde{Y}$ into Eq. \ref{eom}, we get, after some algebra separate dispersion relations for the transverse and longitudinal parts, because we take up to second order $\pi^I$.  
\begin{equation}\label{eqvorticity}
\bigg\{  \omega_T^4 - k^2 \omega^2_T \bigg\}   \bigg( \frac{4 c \chi^2(b_o,0)}{b_o F^\prime(b_o)(1+ i\omega_Y\tau_Y)^2} \bigg) - \omega_T^2=0 
\end{equation}
\begin{equation}\label{eqlongitudinal}
\bigg\{ \omega_L^4 + k^2 \omega^2_L \bigg\}  \bigg( \frac{4 c \chi^2(b_o,0)}{b_o F^\prime(b_o)(1+ i\omega_Y\tau_Y)^2} \bigg) - \omega_L^2 + c_s^2 k^2 = 0    
\end{equation}
We can then express $w_Y(w_L,w_T)$ and solve these equations for the group velocity $ v_g= dw_{T,L}/dk$ of the longitudinal and transverse modes.   Unless $0<v_g<1$ for all perturbations, a theory cannot be causal.  Thus, we use a calculation analogous to \cite{romdev,koide} to test for acausal modes.
The dispersion relations are shown in Fig. \ref{dispfig}, where for brevity we
defined with
\[\   B = 4 c \chi^2(b_o,0) \eqcomma A= b_o F^\prime(b_o)   \]
As can be seen, when $\frac{\tau^2_Y }{(B/A)} \simeq 3$ the group velocity is not casual and its asymptotic velocity goes to negative values as we can note in fig 2 for sound modes. 
In the large $k$ limit dispersion relations are monotonic.  In this UV limit the group velocity is calculable analytically.  As this is the limit where deviations from the EFT should manifest themselves, examining it in a bottom-up approach will tell us if the ideal hydrodynamic limit to an arbitrary scale is well-defined. For the ferrovortetic $c> 0$ and anti-ferrovortetic $c<0$ cases we get
\begin{equation}
 \left.  \lim_{ k \gg 1  } \frac{ d | \omega_T |}{d k}   \right|_{c \lessgtr 0} = \frac{1}{\sqrt{ 1 \mp \frac{\tau^2_Y }{(B/A)} }}
\end{equation}
The equivalent for the longitudinal case are
\begin{equation}
\left. \lim_{ k \gg 1  }  \frac{ d | \omega_L |}{d k}  \right|_{c \lessgtr 0} = \sqrt{ \frac{c^2_s \tau_Y^2 \mp (B/A)}{\tau_Y^2 \pm (B/A)}} 
\end{equation}

\begin{figure}[h]
\centering
\includegraphics[height=0.25\textheight]{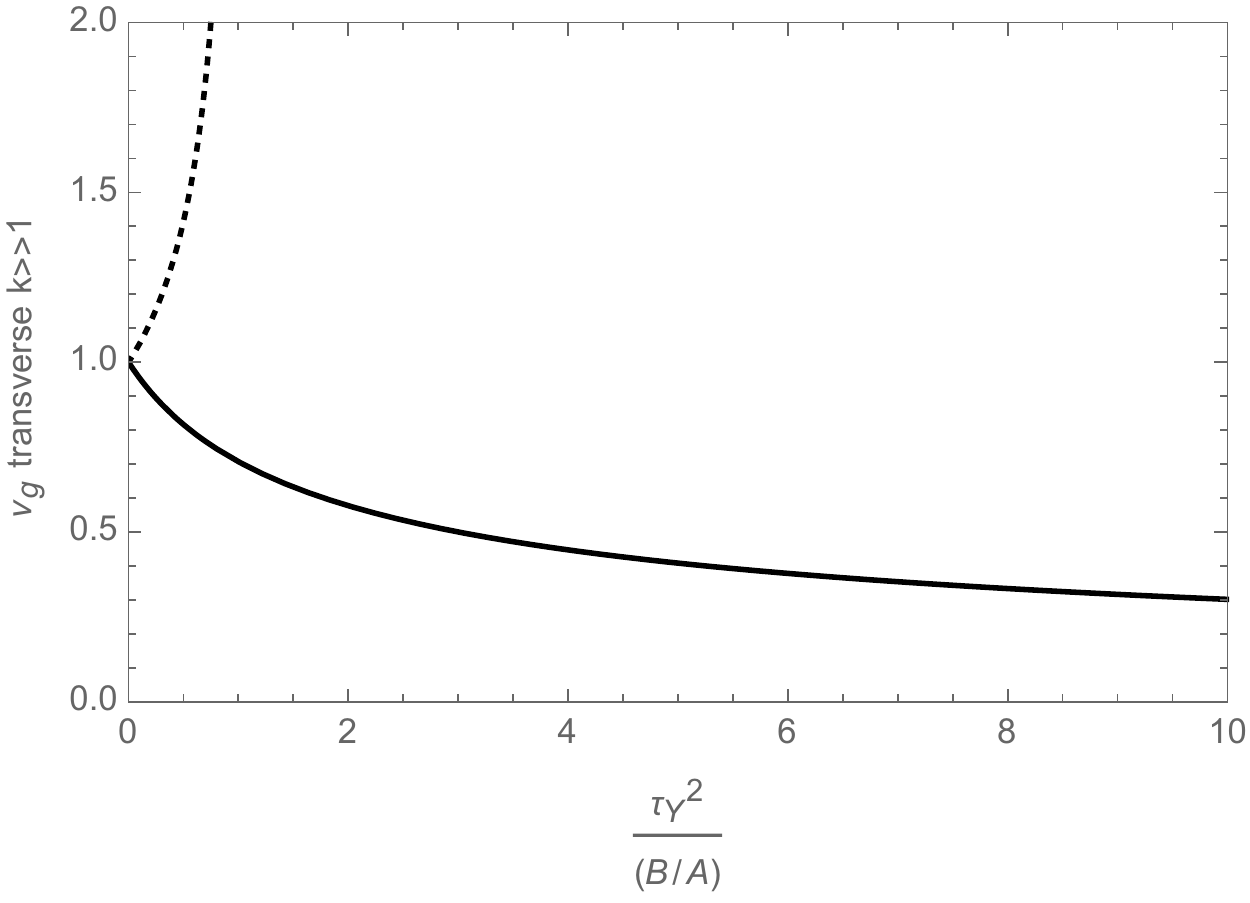}
\includegraphics[height=0.25\textheight]{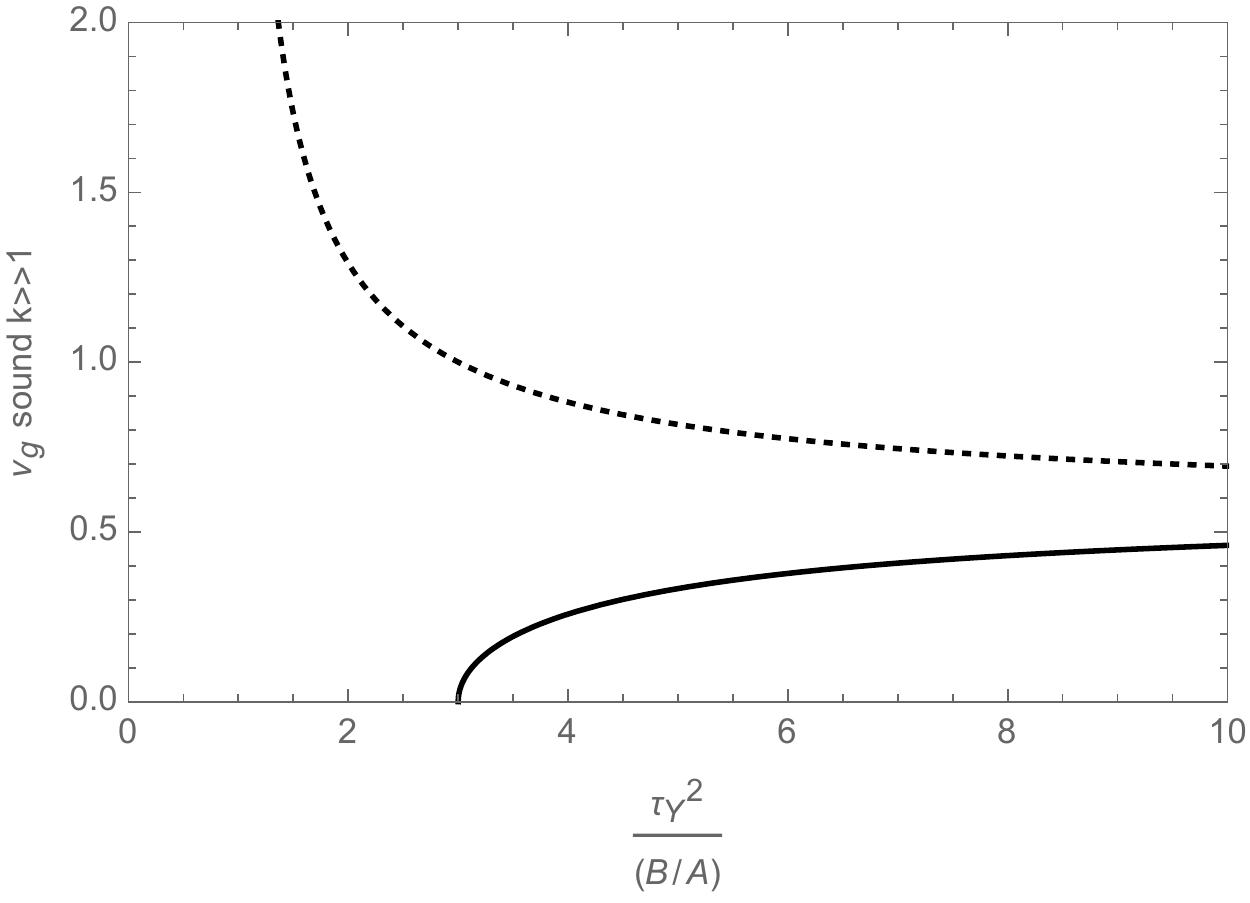}
\caption{\label{calcfigs} The asymptotic value of the group velocities for the transverse and longitudinal modes as a function of the relevant parameter, full and dotted line are $c>0$ and $c<0$, respectively.}
\end{figure}

These are plotted in Fig. \ref{calcfigs}, again for the transverse and longitudinal modes for both ferrovortetic and antiferrovortetic materials.
\section{Discussion\label{disc}}
\subsection{Antiferrovortetic materials and a lower limit on viscosity \label{discferro}}
As can be seen from Fig. \ref{calcfigs}, an antiferrovortetic material can be causal given a constraint on $\tau_Y^2$, given by
\begin{equation}
\label{ineq}
  \tau^2_Y \geq \frac{8  c \chi^2(b_o,0) }{(1-c_s^2) b_o F^{\prime}(b_o)}
\end{equation}
It relates the vortical susceptibility $\chi$ to non-vortical coefficients (speed of sound, enthalphy, hydrostatic entropy).
   The denominator expression $(1-c_s^2) b_o F^{\prime}(b_o)$ is equivalent to $d p/db$ in the polarization-less limit. The numerator is proportional to vorticity's absorption by angular momentum.  Thus, it has exactly the form required of a coefficient describing an effective viscosity arising from spin.   For an unpolarizeable medium (where $\chi=0$ by definition) the lower limit of $\tau^2_Y$ goes to zero, as expected. 
What this shows is that when polarization is present, taking the ultraviolet cutoff $\sim \tau^{-1}_Y$ of hydrodynamic applicability, with zero polarization susceptibility and finite entropy density $b_o$ and $F^\prime (b_0)$, is incompatible with causality.
   
  It should be noted that while this is a relaxation time, its effect is very similar to a viscosity.
  This can be seen by evolving a small vortex with a finite dissipation time.  If the system contains very little vorticity, Eq. \ref{IS} and Eq. \ref{eom} together with a thermodynamically sensible form of $\chi$ ($\chi (|\omega|\rightarrow 0) \rightarrow 0$, as do all its derivatives).   
 The best way to show this for the general case is a Green's function calculation, done in Appendix \ref{app}.  Generally, for a causal medium it can be seen that
 \begin{equation}
\label{expomega}
   \omega_{\mu \nu}(t) \sim \omega_{\mu \nu}(t=0) \exp \left[ -\frac{t}{\tau_Y}\right]
   \end{equation}
 Such an evolution corresponds to the expectation from the definition of a Kubo formula such as Eq. \ref{tauydef}.  
 
Thus, vortex fluid perturbations dissipate into microscopic spin angular momentum and heat on a timescale $\tau_Y$.   But, as is apparent already from the discussion in section \ref{bottomup} and the definition Eq. \ref{vortdef}, vortices in a viscous medium dissipate on a timescale $\eta/(sT)$ \cite{jeon}.   Putting these two scales together we get that the viscosity $\eta$ over entropy density $s$ is bounded by
\begin{equation}
\label{visclimit}
  \eta/s \geq T \tau_Y^{lim}.
  \end{equation}
with $\tau_Y^{lim}$ saturating Eq. \ref{ineq}.
This constraint again makes sense, since the right hand side $\sim T\times \chi$.   In a system with a large degeneracy for a finite amount of energy $c T\chi(b,0) \rightarrow 0$, hence the limit of $\eta/s$ argued for here goes to zero.  
\subsubsection{Some considerations on a bottom-up lower limit on viscosity \label{bottomup}}
The result illustrated above connects to a question which is much more general
and profound than the problem directly dealt with in this work.

The question of weather there exists a universal limit to viscosity and/or dissipation (parameterized, in relativistic systems by the viscosity over entropy ratio $\eta/s$) is both profound and difficult to handle.  On a fundamental level, it is plausible to argue that quantum uncertainty gives rise to fluctuations which dissipate information.
However, translating this realization into a ``bottom up'' limit, independent of a microscopic theory, is problematic.   From the fundamental point of view, moreover, it is unclear ''where the dissipation even comes from'' since the quantum mechanical evolution  is non-dissipative even if the initial state is initially infinitesimally close to equilibrium.  Generally \cite{tong} one assumes that the system is open and coupled to a thermal bath, with the coupling modeled as a time-dependent source. Dissipation coefficients can then be extracted from the response.

In quantum field theory, where the number of degrees of freedom is continuously infinite, we can assume an infinitesimal departure from thermal equilibrium in a thermodynamic limit and calculate the infrared (long frequency-wavenumber limit) response, which is dissipative \cite{tong} provided the vacuum is stable.  
  This is the principle under which we proposed Eq. \ref{tauydef}, as all transport coefficients are calculated similarly.

Thus, given a microscopic theory, the shear viscosity $\eta$ is then generally related \cite{jeon,mooresound} to the correlator of the off-diagonal components of the energy-momentum
tensor $T_{ij,i\ne j}$ via Kubo's formula, defined in Euclidean space as
\begin{equation}
\label{kubo}
\eta = \lim_{w,k \rightarrow 0} w^{-1} \mathrm{Im} \int  \theta (x^0) \ave{ \left[ T_{ij}(x) T_{ij}(0) \right]} e^{  i(k x- w x^0) }dx dx^0
\end{equation}
the real part of the correlator is $w\times p$ \cite{jeon,mooresound}, and hence can be used to obtain the entropy density via thermodynamic identities $s=dp/dT$.

This allows us in principle to calculate transport coefficients given a thermally equilibrated microscopic theory which is also tractable.
However, since relativistic systems with low viscosity are usually strongly coupled, this is a very blunt instrument for claiming ``universal'' limits since Equation \ref{kubo} can not be calculated even approximately in the strongly coupled limit.   

A fundamental limit has been claimed heuristically decades ago by combining the uncertainty principle with Boltzmann's derivation of viscosity \cite{dangyul}, $\eta/s \sim \order{0.1}$.  While this is a plausible order-of-magnitude estimate, it was always clear that Boltzmann's derivation should not generally apply to strongly coupled quantum fields because microscopic correlations, in a strongly coupled system, will ensure that all $n-$point functions will contribute equally to Equation \ref{kubo} so the Boltzmann equation (which only keeps 1-point functions) is inadequate.

More recently, Gauge-gravity correspondence allowed to conclude \cite{kss} that theories with a classical gravity dual have $\eta/s=(4\pi)^{-1}$ in their strong-coupling limit.   The universality of this limit is a consequence of the black-hole no-hair theorem, and hence it critically depends on the existence of a classical gravity dual, namely a planar limit (and consequently an infinite heat capacity) and a conformal strongly coupled fixed point.  Counter examples have been argued for beyond this limit \cite{buchel,cohen,llanes}, which makes its relevance to systems well away from the planar limit dubious.

These difficulties illustrate that most likely one cannot get a lower limit from top-down arguments, where hydrodynamics appears as a limit of a known microscopic theory, and this theory is used to calculate the right-hand side of Eq. \ref{kubo}.
A bottom-up constraint, based on effective field theory constraints such as low-energy unitarity and causality of the fluid dynamics, is necessary.  Such a constraint, if it exists, would imply that any consistent and causal theory would have an $\eta$ or $\eta/s$ above a certain value.

Attempts in this direction can be formulated in terms of the necessity to renormalize the Kubo formulae with hydrodynamic fluctuations \cite{mooresound,kovtun,mauricio}, energy conditions \cite{hartnoll}, the necessity of a quantum cutoff for the Kolmogorov cascade \cite{gthydro} and other arguments, typically related to the number of microscopic degrees of freedom per unit volume (infinite in the ``planar limit'' but finite for a realistic theory), rather than the Knudsen number/gradient expansion.

We note that this planar limit can be thought of as a ``$(thermodynamic)^2$'' limit, where not just the total number of degrees of freedom, but the number of degrees of freedom per unit volume diverges\footnote{Note that some models have been constructed where it is claimed only entropy density diverges, invalidating any bounds on $\eta/s$ \cite{cohen,llanes}}.   This can be seen explicitly in a perturbative calculation \cite{gthydro} of a deformed ideal hydrodynamic limit \cite{nicolis1}, where a finite $\eta/s$ arises only for a diverging microscopic degeneracy.
Thus, it is a deviation from this limit (obviously unrealistic since $N_c=3 \ll \infty$, and, given the applicability of hydrodynamics for systems with $<100 fm^{-3}$ degrees of freedom, most likely not a good approximation) that a bottom-up limit from viscosity might turn up.
However, as vortex degrees of freedom in three dimensions appear strongly non-perturbative \cite{nicolis1,nicolis3,gripaios}, analytical progress in this direction is not easy.

So far, the most quantitative argument we have relating the mean free path to the viscosity is to assume the microscopic scale to be an ultraviolet cutoff and calculating loop corrections \cite{mooresound,kovtun,mauricio}.  The result of such calculations is plausible, but the non-perturbative dynamics of vortices as well as the fact that the UV cut-off is imposed by hand means one can not think of it as fully established.
A promising direction for its completion could be to link the existence of the cutoff to causality, following the link found in \cite{hartnoll} to the null energy condition.  

The question then is, can Eq. \ref{visclimit} be interpreted in such a way?

While polarization appears irrelevant to viscosity, all known physical realizations of strongly coupled fluids as well as most non-trivial interacting field theories contain particles with spin, and the strongly coupled dynamics of such systems must self-evidently include spin-orbit interactions.

So, can the limit found in the previous section be a candidate for such a ``bottom-up viscosity limit?''
Comparing Eq \ref{kubo} to the result of the appendix (Eq. \ref{expomega}), it can be seen that indeed $\tau_Y$ behaves exactly as $\eta/(sT)$ as shown in Eq. \ref{visclimit}.   Qualitatively, transverse modes ``dissipate`` into polarization, and longitudinal and transverse modes are inherently mixed by finite susceptibility.

The behavior of $\tau_Y$ with number of degrees of freedom  is also consistent with the expectation that such ``fluctuation-like`` effects go away in the limit of ``many degrees of freedom'' per unit volume, the ``square of the thermodynamic limit'' \cite{largen}.  In this limit (corresponding to the planar limit in Yang-Mills theories and the applicability of molecular chaos in transport equations), the amount of angular momentum redistributed in polarization DoFs ($\sim \chi$) vanishes.  Thus, the bound found here is "orthogonal" to the celebrated bound of \cite{kss}, and is relevant for systems which are strongly coupled but with comparatively few degrees of freedom per unit volume.   The Quark-Gluon Plasma and ultracold atoms would be obvious examples.  

The limit in Eq. \ref{visclimit} is ``bottom-up'', inasmuch as it is only dependent on the assumption of causality, local thermalization and symmetries and that, unlike Israel-Stewart, it corrects an ideal fluid limit rather than one which is already dissipative.  $\tau_Y$ as a cutoff is generated dynamically by the fundamental quantization of spin, rather than imposed by hand in loop corrections.
This cutoff also breaks up turbulence cascades which make vortices unstable \cite{nicolis1}, as expected \cite{gthydro} from a "quantum" (because of the presence of spin) viscosity limit.  \begin{figure}[h]
\includegraphics[height=0.12\textheight]{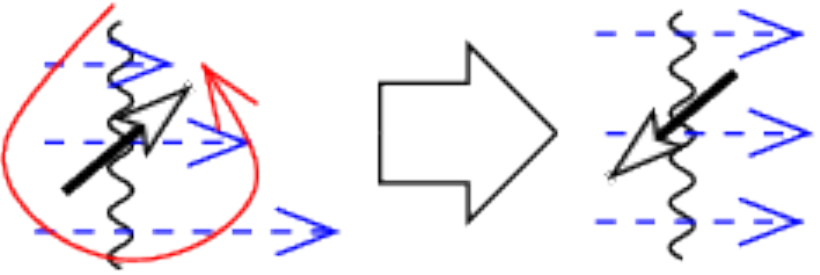}
\caption{\label{quali}An illustration of how polarization of a quantum particle could lead to an effective viscosity independent of the mean free path  }
\end{figure}

The fundamental issue of how to reconcile the loss of unitarity inherent in dissipation with the unitary nature of quantum evolution is of course still mysterious on a rigorous level.  However, there are a few things we can say.
 While the theory used here is classical, it follows from the intrinsically quantum notion of ``spin'' of a point particle \cite{goudsmit} and any qualitative explanation requires a finite de Broglie wavelength.

Qualitatively, the effect we derived can be understood from a microscopic quasi-particle picture, as illustrated in Fig \ref{quali}:  A particle with spin and a finite de-Broglie wavelength \cite{gauge}, moving in a fluid with momentum flow, could, given any spin-orbit interaction, have its helicity flipped by the gradient of the momentum density, which in a flowing fluid is the gradient of the flow.  By angular momentum conservation, the helicity flip will quench some of the gradient (note that in relativistic fluids vorticity is defined via the energy momentum tensor used in Eq.\ref{kubo} \cite{flork1,flork2}), thereby ensuring that gradients of macroscopic quantities impact parameters of the distribution of microscopic particles. Given Stokes theorem and the definition of viscosities via Eq. \ref{kubo}, these helicity-momentum interactions will have the same effect as a shear viscosity.  Note that, in the strong limit of the spin-orbit coupling this effect does not go away, only for short de Broglie wavelengths (i.e. high temperature) or many microscopic degrees of freedom (the planar limit in Yang-Mills theories) it will disappear, just like the dissipative effects of microscopic fluctuations\cite{gthydro,mooresound}.   
That this effective viscosity is dissipative can be realized by an entropy analysis \cite{hongo}, since entropy will invariably depend on flow gradients.
Thus, in the thermodynamic limit but with a finite number of degrees of freedom per unit volume, quantum uncertainties combined with the quantum internal structure of each degree of freedom can result in an effect mimicking viscosity which, in a field theory setting, can give rise to a finite $\eta/s$.
\subsection{Ferrovortetic materials. Transition to a polarized phase? \label{discanti}}
For the ferrovortetic ($c>0$) material an acausal mode remains, and, as one can see from Fig. \ref{dispfig} this mode is infrared (small $k$) rather than ultraviolet (large $k$).    
Note that the longitudinal mode only diverges at a critical $k$ while the
transverse mode is always acausal.   A simple explanation would be to suppose
$c>0$ is unphysical, but then one would not understand why this is so, spontaneous spin
alignement is known to microscopically occur in the non-relativistic limit.

There is however a physically compelling hypothesis, that the infrared mode is rather related to the thermodynamic vacuum instability of the system.
The unpolarized vacuum is a false vacuum, and the decay into a true vacuum can be seen as an infinitely soft perturbation.
Within the context of chiral phase transitions this is known as a Banks-Casher \cite{casher} mode, relating the spectral function at $k \lim 0$ to the appearance of the condensate (in this case of spin alignement rather than chiral) and the spectral function $\rho(w)$.
\begin{equation}
\label{bk}
 \ave{y_{\mu \nu}} \equiv \lim_{k \rightarrow 0} \rho(\omega_{T,L}(k) )
\end{equation}
where $\rho(\omega)$ is the spectral function.   Such a relation however is relevant to any relativistic system where a condensate, in this case a spin condensate, appears. Physically, what happens is that a low wavelength mode is indistinguishable from the formation of a spin condensate, and indeed below a critical temperature such a formation is unavoidable, and hence the ``violation of causality'' signals the appearance of spontaneous polarization.
   
As expected from fluctuation-dissipation arguments, the wavenumber $k$ (Fig. \ref{dispfig}) where causality breaks down is related to the size of the Domain wall where fluctuations and spontaneous spin alignement are comparable.  Locally, the instability under spin alignement means that
the hydrostatic vacuum is {\em always} unstable under vortex formation, so $\pi_T$ is always acausal in the linear order.

In this case Eq. \ref{IS} is not anymore a good effective theory, since the fluid degrees of freedom and the spin condensate will evolve and interact with their own equations of motion.  Hence, Eq. \ref{bk} cannot be used to calculate the condensate.   In fact, the vacuum instability signals that the expansion of Eq. \ref{lagdef} needs to be augmented, at least, with the quartic term
\begin{equation}
   F(b,y) = F\left(b \left(1-c(T,\omega)\, y_{\mu\nu}y^{\mu\nu}+ c_4(T,\omega) \order{y^4}\right)\right), 
\end{equation}
This equation is exactly that of the free energy for Landau phase transitions, and indeed in general there will be a condensate defined by $\ave{y} \ne 0$ and the minimization of $F(b,y)$, in exact analogy with ferromagnetism.

However, we are considering a locally equilibrated fluid here, including thermal fluctuations and sound-waves.   These conditions should generate different condensate domains, where thermal fluctuations and sound waves are in competition with spontaneous condensation.
all these effects could be found by
calculating the functional integral
\begin{equation}
\label{effz}
\ln Z= \int \mathcal{D} \left[y,\phi_I \right] e^{T_0^4 \int \mathcal{L}(y,\phi_I,c) d^4 x} \simeq L_{eff}\left[y',\phi_I',c'\right]_{T_0}
\end{equation}
in terms of the microscopic scale $T_0$ \cite{gthydro,ushydro,burch}.
Using the lagrangian given here both fluid fluctuations and polarization will be treated on the same footing and could give the interplay between spontaneous polarization, thermal fluctuations and hydrodynamic evolution which would manifest in a renormalization group flow of $c$ between a polarized and a depolarized phase defined in terms of an effective action \cite{peskin}
\begin{equation}
  \label{gammadef}
 \Gamma (\ave{y}) = \ln Z - \int d^4 z J^{\mu \nu}(z) \ave{y_{\mu \nu}}
  \end{equation}
We remark that, since this is correcting an ideal locally equilibrated fluid, the time-scale $\tau_Y$ and the gradient around the local minimum of Eq. \ref{gammadef} should be related to ensure an equilibrium, as per the usual fluctuation dissipation theorem and the Kramers-Koenig relations \cite{tong}.

Analyzing this in detail is a very ambitious, possibly numerical project that shall be left for further work.
\subsection{Discussion and conclusions}
This work had two motivations:  The specific issue with causality in the ideal hydrodynamic limit when the fluid has a non-zero polarization, i.e. when its microscopic constituents have a non-zero spin.   As shown in \cite{polhydro2}, the ideal limit is incompatible with causality.  In this work, we have shown that an Israel-Stewart like approach can fix this problem.  However, unlike with Israel-Stewart hydrodynamics, the limit theory is ideal.  Hence, rather than correct a dissipative theory we need to introduce dissipation in a non-dissipative one.

This makes a connection to a much more general issue: The possible existence of a ``bottom up'' lower limit on viscosity for general fluid-like systems, which does not depend on specific assumptions on the microscopic theory governing the constituents of the fluid.   The minimal dissipation we find can be thought of as just such a limit, since the dynamics of this dissipation mimics that of viscosity.  

We also showed the different behavior of the ferrovortetic limit (where polarization lowers free energy) from the antiferrovortetic one (where it increases the free energy).
It is in the ultraviolet limit of the antiferrovortetic case that the non-causal mode is damped by the Israel-Stewart dissipation, just as expected for a viscous correction.
In the ferrovortetic regime non-causality appears, but in the infrared rather than the ultraviolet limit.  Its physical interpretation is different, signaling the instability of the fluid against spontaneous polarization.

We close with some phenomenological considerations.    To test this theory experimentally one must be able to convert our polarized fluid to particles.
The usual method to do so is via the Cooper-Frye formula \cite{cf} and its viscous extensions.  The Cooper-Frye formula has already been extended to include conversion of angular momentum into spin \cite{bec1} at freezeout.
To link with our work we need to understand how freezeout happens when both a vorticity and a spin current $Y_{\mu \nu}$ exist when the fluid freezes out into particles.

The problem is that the Cooper-Frye formula is based only on conservation laws and entropy non-decrease  at freeze-out.   This, as shown in \cite{polhydro1} is {\em not} enough to define dynamics for a fluid with polarization.  Hence, more dynamics will have to be added to the Cooper-Frye formula once spin and angular momentum are separately handled.

One physically appealing way to do it is to use the Wigner function formalism for coalescence.   $Y_{\mu \nu}$ could be straight-forwardly linked to the spin wave-function of the constituent quarks/gluons (One would attach an Eigenstate to each coefficient instead of the generator \blacktext{ in Eq. \ref{generat}}) and vorticity could be linked to the angular momentum part of the wavefunction.  Since this process is quantum, entropy would be conserved.   However, in practice, a lot of untested assumptions (how many constituent quarks per each hadron?) would have to go into such a model before a meaningful connection with data is made.

The phenomenological manifestation of pre-existing spin is also non-trivial to investigate via experimental observables.  As shown in \cite{karpenko} \blacktext{and \cite{csernai}}, to describe transverse $\Lambda$ polarization it is enough to assume vorticity is transferred to spin only at freezeout.
So, unsurprisingly, the effects studied in this paper might well be sub-leading.   Longitudinal polarization \cite{longi} was explained within non-equilibrium dynamics incorporating thermal vorticity \cite{ampt}.   Perhaps comparing polarization of $\Lambda$ and vector mesons \cite{align1,align2} might show the need to go beyond the Cooper-Frye approach, while electromagnetic probes \cite{emvort} will show the necessity to propagate polarization as well as vorticity from the hot initial phase where flow anisotropies develop.

For phenomenological comparisons, one would also need to link with microscopic physics.   To date, a lattice calculation of QCD in a rotating frame is available \cite{rotlat} and calculations in effective theories are ongoing \cite{spinal}.  In principle one can use them to calculate $c$, distinguish between a ferrovortetic and an antiferrovortetic case, and compute $\chi$ and $\tau_Y$ via Eq. \ref{tauydef}, but as yet we do not know if QCD matter is ferrovortetic, antiferrovortetic or both at different temperatures.

Hence, a tight phenomenological test of this theory is still far away.
However, 
the fact that Lagrangian hydrodynamics can capture both textbook physics (spontaneous spin condensation) and a widely expected but never quite proven lower limit on dissipation in strongly coupled systems certifies its status as a powerful theoretical tool to examine the behavior of relativistic fluids.
The Lagrangian proposed in Eq. \ref{genlag} can therefore be considered as a candidate for the Lagrangian of a polarizeable medium close to the ideal fluid limit.   A connection of this hydrodynamics to both microscopic theories and phenomenology will be studied in forthcoming work.

\textit{Acknowledgments}   GT acknowledges support from FAPESP proc. 2017/06508-7, 
partecipation in FAPESP tematico 2017/05685-2 and CNPQ bolsa de
 produtividade 301996/2014-8. DM was supported by a CAPES graduate fellowship. This work is a part
of the project INCT-FNA Proc. No. 464898/2014-5.  We thank Radoslaw Ryblewski,Mike Lisa and Nandini Trivedi for fruitful discussions and suggestions.

\appendix*
\section{Evolution of a vortex in a thermal polarizeable causal medium \label{app}}
 The best way to approach this problem is to unfold Eq. \ref{eqvorticity} into configuration space.   After a tedius but straight-forward calculation we have
\textcolor{black}{
\begin{flalign*}
B \frac{\partial^4}{\partial t^4}  \pi_T(\vec{x},t)   - A \frac{\partial^2}{\partial x^2} \frac{\partial^2}{\partial t^2}  \pi_T(\vec{x},t) + &&
\end{flalign*}
\begin{equation}\label{123}
  2 \tau \frac{\partial^3}{\partial t^3}  \pi_T(\vec{x},t)
+ \frac{\partial^2}{\partial t^2} \pi_T(\vec{x},t) = 0
\end{equation}
where the constant variables are 
\begin{equation}
B = A + \tau^2, \qquad  A = \frac{4 c \chi^2(b_o,0)}{b_0 F^{\prime}(b_o)}
\end{equation}}

One simple redefinition of main variable $\frac{\partial^2}{\partial t^2} \pi_T(\vec{x},t) = \Pi_T(\vec{x},t) = X(x)T(t)$  allow us solve by  separation of variable method. Assuming $ \pi_T(\vec{x},t)$ is a analytical function and infinitely differentiable Eq. \ref{123} can be rewritten as

\textcolor{black}{
\begin{flalign}
  \frac{1}{T(t)} \bigg( B \frac{\partial^2}{\partial t^2} T(t) + 2 \tau \frac{\partial }{\partial t } T(t) \bigg) = &&
  \nonumber
\end{flalign}
\begin{equation}
 \frac{1}{X(x)}  \bigg( A \frac{\partial^2}{\partial x^2} X(x) - X(x) \bigg)  = - \lambda
\end{equation}}

To construct the general solution, first, we seek solve   each part separately: spatial and time., and afterwards Substituting into $\Pi_T(\vec{x}, t)$.

\subsubsection{Spatial Part}

Turning now to spatial differential equation that will be solved as ordinary one.

\begin{equation}
  \bigg( A \frac{\partial^2}{\partial x^2} X(x) - ( 1  - \lambda) X(x)   \bigg)  = 0
\end{equation}

By characteristic method, the roots of polynomial equation will be the part of general homogeneous solution.  
\textcolor{black}{\begin{equation}
X(\vec{x})  = \gamma_1(t) e^{ ix \chi^{-1} \sqrt{\frac{(1-\lambda)}{A'} }} + \gamma_2(t) e^{- ix \chi^{-1} \sqrt{\frac{(1-\lambda)}{A'}}}
\end{equation}}

with $\gamma_1$ and $\gamma_2$ are smooth functions which may depend on time, with the solution being unique in any interval where $\gamma_{1,2}$ is continuus.
At first, such an ''oscillating vortex'' is unexpected, but a ''wavepacket'' of such solutions will give a familiar localized vortex.

We can always expand on one complete closed set of orthogonal function \textcolor{black}{since the operator is linear and self-adjoint, and we can expand Green function in terms of this set}. Being an invertible operator is straightforward to evaluate a response function
\textcolor{black}{\[\   G(\vec{x}^{\prime} - \vec{x}) = - \mathcal{L}^{-1} \delta^3(\vec{x}^{\prime} - \vec{x})
\]}
In our particular case
\begin{equation}
G_T(\vec{x}^{\prime} - \vec{x}, t^{\prime} - t ) = \int \frac{d^3 k}{(2\pi)^3} e^{-i \vec{k}(\vec{x}^{\prime} - \vec{x})} G_T(\vec{k}, t^{\prime} - t)     
\end{equation}
  \[\   G_T(k, t^{\prime} - t)= \frac{1}{k^2 - \frac{(1-\lambda)}{2 A} + i\epsilon}   \]
  This greens function evolves only the vorticity term $\pi_T$, which is perpendicular to momentum.
As such, it has no isotropy in frequency space, but rather an anisotropy due to thermalization process that correspond to angular velocity.  THe Greens function evolving $\pi_L$ will be different, and will be examined in a forthcoming work.
\subsubsection{Time evolution}
The evolution of the time dependent part of the equation
\begin{equation}
 \bigg(  \frac{\partial^2}{\partial t^2} T(t) + \frac{2 \tau}{B} \frac{\partial }{\partial t } T(t)   - \frac{\lambda}{B} T(t) \bigg) = 0
\end{equation} 
is similar to a damped harmonic oscillator. By the method of characteristics
we have, after defining
\begin{equation}
r = - \frac{ \tau }{B} \pm \sqrt{\frac{\tau^2}{B^2} - \frac{\lambda}{B}  }
\end{equation}
\textcolor{black}{\begin{flalign}
\label{ineqdef}
B = (\frac{4}{9}\tau^2_{limit}+ \tau^2 ), \ \  \tau^2_{limit}= \frac{8  c \chi^2(b_o,0) }{(1-c_s^2) b_o F^{\prime}(b_o)}=\frac{9}{4}A
\end{flalign}}

where the speed of sound squared $c_s^2 = 1/3$, and $\frac{9}{4}$ arises after taking hydrodynamic limit on group velocity. The case of interest from three general possible solution 
\textcolor{black}{\begin{description}
\item[Null] $r=0$ Critical Damping
  \[\  \tau^2 = \lambda B \ \ \rightarrow \ \ \tau^2 = \bigg( \frac{\lambda}{1 - \lambda}  \bigg) \frac{4}{9} \tau^2_{limit}\]
  and
  \[\ T(t) =( \alpha(\vec{x})  +\beta(\vec{x}) t ) e^{ - \frac{\tau}{B}t}\]
\item[Real] $r>0$  Overdamping
  \[\
 \tau^2 > \lambda B \ \ \rightarrow \ \ \tau^2 > \bigg( \frac{\lambda}{1 - \lambda}  \bigg) \frac{4}{9} \tau^2_{limit}
 \]
  and
\[\ T(t) = e^{- t \frac{\tau}{B}}( \alpha(\vec{x})e^{\omega_2 t}  +  \beta(\vec{x}) e^{- \omega_2 t}), \quad \omega_2 = \sqrt{\frac{\tau^2 }{\lambda B} - \frac{\lambda}{B}  }
  \]
\item[Imaginary] $r<0$     Underdamping
  \[\
 \tau^2 < \lambda B \ \ \rightarrow \  \ \tau^2 < \bigg( \frac{\lambda}{1 - \lambda}  \bigg) \frac{4}{9} \tau^2_{limit}    
  \]
  and
  \[\  T(t) =  e^{- t \frac{\tau}{B}} e^{i t \sqrt{\frac{\lambda}{B} } - \frac{\tau^2 }{B^2} }  +\beta(\vec{x}) e^{- t \frac{ \tau }{B}} e^{- it \sqrt{\frac{\lambda}{B} -  \frac{\tau^2 }{B^2} }   } \]
\end{description}}

Since $0<\lambda<1$, overdamping corresponds to the causality limit defined by Eq. \ref{ineq}.
The green function is 
\begin{equation}
G_T(x^{\prime}-x, \omega) = \frac{1}{- \omega^2 + i (2\tau/B)\omega + \omega^2_o},
\end{equation}
where

\textcolor{black}{\[\   \omega_{1,2} = - i \tau \pm \sqrt{\omega_o^2 - \gamma^2}  \]}

and
\begin{equation}
 Im G_T(x^{\prime}-x, \omega)  = \frac{(2\tau/B)\omega}{(\omega^2_0 - \omega^2)^2 + (2\tau/B)^2 \omega^2 }
\end{equation}
 \begin{equation}
 Re G_T(x^{\prime}-x, \omega)   = \frac{\omega^2_o - \omega^2}{(\omega^2_0 - \omega^2)^2 + (2\tau/B)^2 \omega^2 }, 
\end{equation}
the damping gradually dissipates initial energy, and only imaginary case results in oscillatory movement. 

The qualitative behavior of the solutions at asymptotic time is thus

\textcolor{black}{\begin{description}
\item[null] its solutions at $t \rightarrow \infty$ tends to zero with a relaxation time $\tau = \frac{2}{3} \tau_{limit} $. The system not even reaches crossing time axis. 
\item[real] Vector perturbation field goes to zero at infinity time slower than critical damping solution. Where boundary constraints is automatically satisfied, $\pi_T(\vec{x},t)$ sets up as dissipative solution under a characteristic time scale. 
\item[imaginary] (Nonphysical) Unstable oscillatory movement across time axis lies between boundary curves $\pm \alpha e^{-t \frac{\tau}{B}}$. Since a damping term is present, we cannot define a frequency in physical meaning of word.  
\end{description}}

Equation \ref{ineq} imposes the imaginary case, which we shall use to obtain the general solution.
\subsubsection{Solution}

\textcolor{black}{
Putting everything together, and ensuring the causality constaint $0 < \lambda < 1$ we get the general evolution equation for $\pi_T$}
\textcolor{black}{
\begin{flalign*}
\pi_T(\vec{x},t) = e^{- \frac{\tau}{B}t}\bigg[\frac{\alpha(\vec{x}) }{(\frac{\tau}{B} + \omega_2)^2  } e^{\omega_2 t} + \frac{ \beta(\vec{x}) }{(\frac{\tau}{B} - \omega_2)^2  } e^{- \omega_2 t}\bigg]\times 
\end{flalign*}
\[\ \times  \bigg[ 
      \gamma_1(t) e^{ ix \chi^{-1} \sqrt{\frac{(1-\lambda)}{A'} }} 
      + \  \gamma_2(t) e^{- ix \chi^{-1} \sqrt{\frac{( 1 - \lambda)}{A'}}} \bigg]+
\]
\[\
+c_1(\vec{x}) t + c_2(\vec{x})
\]
}
where the boundary conditions (existence of equilibrium) imply $c_{1,2}=0$ (the initial background value which would represent $c_1$ is absorbed into $\gamma_{1,2}$ and $c_2$ into their gradients.
This establishes the asymptotic behavior of Eq. \ref{expomega}.



\begin{thebibliography}{99}

\bibitem{polhydro1} 
  D.~Montenegro, L.~Tinti and G.~Torrieri,
  Phys.\ Rev.\ D {\bf 96}, no. 5, 056012 (2017)
  Addendum: [Phys.\ Rev.\ D {\bf 96}, no. 7, 079901 (2017)]
  doi:10.1103/PhysRevD.96.079901, 10.1103/PhysRevD.96.056012
  [arXiv:1701.08263 [hep-th]].

\bibitem{polhydro2} 
  D.~Montenegro, L.~Tinti and G.~Torrieri,
  Phys.\ Rev.\ D {\bf 96}, no. 7, 076016 (2017)
  doi:10.1103/PhysRevD.96.076016
  [arXiv:1703.03079 [hep-th]].

  
\bibitem{bec2}
  F.~Becattini and L.~Tinti,
  Annals Phys.\  {\bf 325}, 1566 (2010)
  [arXiv:0911.0864 [gr-qc]].


\bibitem{gauge}
  Care must be taken here with gauge symmetry, since the separation between transverse spin and the spatial propagator, i.e. the orbital angular momentum should leave the action, and hence the free energy, unchanged.
  For further details, see\\
  G.~Torrieri,
  ``Swimming and swirling colorful ghosts,''
  arXiv:1810.12468 [hep-th].


\bibitem{explanation}  The terminology ''chiral vortaic effect'' is confusing, since it is often used in the context of the chiral magnetic effect.  However, the chiral magnetic effect is charge-dependent and hence anomalous, the chiral vortaic effect is charge-blind and not anomalous, and both come with vector and axial vector terms.
  For a comprehensive review, see\\
  S.~Lin and L.~Yang,
  arXiv:1810.02979 [nucl-th].\\



\bibitem{bec1}
  F.~Becattini, V.~Chandra, L.~Del Zanna and E.~Grossi,
  Annals Phys.\  {\bf 338}, 32 (2013)
  doi:10.1016/j.aop.2013.07.004
  [arXiv:1303.3431 [nucl-th]].

  
\bibitem{shu} 
  Y.~Hidaka, S.~Pu and D.~L.~Yang,
  Phys.\ Rev.\ D {\bf 95}, no. 9, 091901 (2017)
  doi:10.1103/PhysRevD.95.091901
  [arXiv:1612.04630 [hep-th]].

\bibitem{vortsus}
  A.~Aristova, D.~Frenklakh, A.~Gorsky and D.~Kharzeev,
  JHEP {\bf 1610}, 029 (2016)
  doi:10.1007/JHEP10(2016)029
  [arXiv:1606.05882 [hep-ph]].


\bibitem{flork1} 
  W.~Florkowski, A.~Kumar and R.~Ryblewski,
  arXiv:1806.02616 [hep-ph].

\bibitem{flork2} 
  W.~Florkowski, B.~Friman, A.~Jaiswal and E.~Speranza,
  Phys.\ Rev.\ C {\bf 97}, no. 4, 041901 (2018)
  doi:10.1103/PhysRevC.97.041901
  [arXiv:1705.00587 [nucl-th]].

\bibitem{hongo} 
  K.~Hattori, M.~Hongo, X.~G.~Huang, M.~Matsuo and H.~Taya,
  arXiv:1901.06615 [hep-th].
  


\bibitem{nature}
  L.~Adamczyk {\it et al.} [STAR Collaboration],
  Nature {\bf 548}, 62 (2017)
  doi:10.1038/nature23004
  [arXiv:1701.06657 [nucl-ex]].

\bibitem{barnett} 
  K.~Fukushima, S.~Pu and Z.~Qiu,
  Phys.\ Rev.\ A {\bf 99}, no. 3, 032105 (2019)
  doi:10.1103/PhysRevA.99.032105
  [arXiv:1808.08016 [hep-ph]].


\bibitem{nicolis1}
  S.~Endlich, A.~Nicolis, R.~Rattazzi and J.~Wang,
  JHEP {\bf 1104}, 102 (2011)
  [arXiv:1011.6396 [hep-th]].



\bibitem{burch} 
  T.~Burch and G.~Torrieri,
  Phys.\ Rev.\ D {\bf 92}, no. 1, 016009 (2015)
  doi:10.1103/PhysRevD.92.016009
  [arXiv:1502.05421 [hep-lat]].


  \bibitem{nicolis3}
  S.~Dubovsky, L.~Hui, A.~Nicolis and D.~T.~Son,
  arXiv:1107.0731 [hep-th].


\bibitem{gthydro} 
  G.~Torrieri,
  Phys.\ Rev.\ D {\bf 85}, 065006 (2012)
  doi:10.1103/PhysRevD.85.065006
  [arXiv:1112.4086 [hep-th]].

  

\bibitem{ushydro} 
  D.~Montenegro and G.~Torrieri,
  Phys.\ Rev.\ D {\bf 94}, no. 6, 065042 (2016)
  doi:10.1103/PhysRevD.94.065042
  [arXiv:1604.05291 [hep-th]].



\bibitem{gripaios} 
  B.~Gripaios and D.~Sutherland,
  Phys.\ Rev.\ Lett.\  {\bf 114}, no. 7, 071601 (2015)
  doi:10.1103/PhysRevLett.114.071601
  [arXiv:1406.4422 [hep-th]].

  

\bibitem{limostro} 
  T.~j.~Chen, M.~Fasiello, E.~A.~Lim and A.~J.~Tolley,
  JCAP {\bf 1302}, 042 (2013)
  doi:10.1088/1475-7516/2013/02/042
  [arXiv:1209.0583 [hep-th]].


\bibitem{grozdanov}
  S.~Grozdanov and J.~Polonyi,
  Phys.\ Rev.\ D {\bf 91}, no. 10, 105031 (2015)
  doi:10.1103/PhysRevD.91.105031
  [arXiv:1305.3670 [hep-th]].

\bibitem{galley}
  C.~R.~Galley, D.~Tsang and L.~C.~Stein,
  arXiv:1412.3082 [math-ph].





\bibitem{rezzolla} L. Rezzolla and O. Zanotti, ''Relativistic hydrodynamics'', Oxford University Press, (2013)
  

\bibitem{israel} 
  W.~Israel and J.~M.~Stewart,
  Annals Phys.\  {\bf 118}, 341 (1979).
  doi:10.1016/0003-4916(79)90130-1

\bibitem{boulware} 
  S.~Deser and D.~Boulware,
  J.\ Math.\ Phys.\  {\bf 8}, 1468 (1967).
  doi:10.1063/1.1705368
  
  \bibitem{coming}
  D. Montenegro, G. Torrieri ''Kubo formulae for polarizeable media'', to be published


\bibitem{mooretaupi} 
  G.~D.~Moore and K.~A.~Sohrabi,
  Phys.\ Rev.\ Lett.\  {\bf 106}, 122302 (2011)
  doi:10.1103/PhysRevLett.106.122302
  [arXiv:1007.5333 [hep-ph]].
    

\bibitem{jorgeis} 
  F.~S.~Bemfica, M.~M.~Disconzi and J.~Noronha,
  arXiv:1901.06701 [gr-qc].


\bibitem{romdev}
  R.~Baier, P.~Romatschke, D.~T.~Son, A.~O.~Starinets and M.~A.~Stephanov,
  JHEP {\bf 0804}, 100 (2008)
  doi:10.1088/1126-6708/2008/04/100
  [arXiv:0712.2451 [hep-th]].

\bibitem{koide} 
  S.~Pu, T.~Koide and D.~H.~Rischke,
  Phys.\ Rev.\ D {\bf 81}, 114039 (2010)
  doi:10.1103/PhysRevD.81.114039
  [arXiv:0907.3906 [hep-ph]].


\bibitem{tong} D.Tong, lectures on kinetic theory, publically available at\\
{\em http://www.damtp.cam.ac.uk/user/tong/kinetic.html}

   
\bibitem{jeon} 
S. Weinberg, gravitation and cosmology.



\bibitem{mooresound} 
  P.~Kovtun, G.~D.~Moore and P.~Romatschke,
  Phys.\ Rev.\ D {\bf 84}, 025006 (2011)
  doi:10.1103/PhysRevD.84.025006
  [arXiv:1104.1586 [hep-ph]].

\bibitem{dangyul} 
  P.~Danielewicz and M.~Gyulassy,
  Phys.\ Rev.\ D {\bf 31}, 53 (1985).
  doi:10.1103/PhysRevD.31.53


\bibitem{kss}
  G.~Policastro, D.~T.~Son and A.~O.~Starinets,
  Phys.\ Rev.\ Lett.\  {\bf 87}, 081601 (2001)
  [arXiv:hep-th/0104066].


\bibitem{buchel} 
  A.~Buchel, R.~C.~Myers and A.~Sinha,
  JHEP {\bf 0903}, 084 (2009)
  doi:10.1088/1126-6708/2009/03/084
  [arXiv:0812.2521 [hep-th]].


\bibitem{cohen} 
A.~Cherman, T.~D.~Cohen and P.~M.~Hohler,
  JHEP {\bf 0802}, 026 (2008)
  doi:10.1088/1126-6708/2008/02/026
  [arXiv:0708.4201 [hep-th]].\\
  See also 
    D.~T.~Son,
  Phys.\ Rev.\ Lett.\  {\bf 100}, 029101 (2008)
  doi:10.1103/PhysRevLett.100.029101
  [arXiv:0709.4651 [hep-th]].


\bibitem{llanes} 
  A.~Dobado and F.~J.~Llanes-Estrada,
  Eur.\ Phys.\ J.\ C {\bf 51}, 913 (2007)
  doi:10.1140/epjc/s10052-007-0332-5
  [hep-th/0703132].

\bibitem{kovtun} 
  P.~Kovtun,
  J.\ Phys.\ A {\bf 45}, 473001 (2012)
  doi:10.1088/1751-8113/45/47/473001
  [arXiv:1205.5040 [hep-th]].

 
\bibitem{mauricio} 
  M.~Martinez and T.~Schäfer,
  Phys.\ Rev.\ A {\bf 96}, no. 6, 063607 (2017)
  doi:10.1103/PhysRevA.96.063607
  [arXiv:1708.01548 [cond-mat.quant-gas]].



\bibitem{hartnoll} 
  L.~V.~Delacrétaz, T.~Hartman, S.~A.~Hartnoll and A.~Lewkowycz,
  arXiv:1805.04194 [hep-th].




 

\bibitem{largen} 
  B.~Lucini and M.~Panero,
  Phys.\ Rept.\  {\bf 526}, 93 (2013)
  doi:10.1016/j.physrep.2013.01.001
  [arXiv:1210.4997 [hep-th]].

  \bibitem{goudsmit} G.E. Uhlenbeck and S. Goudsmit,  Nature {\bf 117} 264 (1926) 


\bibitem{casher} 
  T.~Banks and A.~Casher,
  Nucl.\ Phys.\ B {\bf 169}, 103 (1980).
  doi:10.1016/0550-3213(80)90255-2


\bibitem{peskin} 
  M.~E.~Peskin and D.~V.~Schroeder,

\bibitem{cf} 
  F.~Cooper and G.~Frye,
  Phys.\ Rev.\ D {\bf 10}, 186 (1974).
  doi:10.1103/PhysRevD.10.186

\bibitem{karpenko} 
  I.~Karpenko and F.~Becattini,
  Eur.\ Phys.\ J.\ C {\bf 77}, no. 4, 213 (2017)
  doi:10.1140/epjc/s10052-017-4765-1
  [arXiv:1610.04717 [nucl-th]].

  \blacktext{
\bibitem{csernai} 
  F.~Becattini, L.~Csernai and D.~J.~Wang,
  Phys.\ Rev.\ C {\bf 88}, no. 3, 034905 (2013)
  Erratum: [Phys.\ Rev.\ C {\bf 93}, no. 6, 069901 (2016)]
  doi:10.1103/PhysRevC.93.069901, 10.1103/PhysRevC.88.034905
  [arXiv:1304.4427 [nucl-th]].
  }
  
\bibitem{longi} 
  J.~Adam {\it et al.} [STAR Collaboration],
  arXiv:1905.11917 [nucl-ex].
  
\bibitem{ampt} 
  Y.~Sun and C.~M.~Ko,
  Phys.\ Rev.\ C {\bf 96}, no. 2, 024906 (2017)
  doi:10.1103/PhysRevC.96.024906
  [arXiv:1706.09467 [nucl-th]].

\bibitem{align1} 
  B.~Mohanty [ALICE Collaboration],
  arXiv:1711.02018 [nucl-ex].

\bibitem{align2} 
  B.~I.~Abelev {\it et al.} [STAR Collaboration],
  Phys.\ Rev.\ C {\bf 77}, 061902 (2008)
  doi:10.1103/PhysRevC.77.061902
  [arXiv:0801.1729 [nucl-ex]].

  \bibitem{emvort} 
  J.~R.~Bhatt, H.~Mishra and B.~Singh,
  Phys.\ Rev.\ D {\bf 100}, no. 1, 014016 (2019)
  doi:10.1103/PhysRevD.100.014016
  [arXiv:1811.08124 [hep-ph]].

\bibitem{rotlat}
  A.~Yamamoto and Y.~Hirono,
  Phys.\ Rev.\ Lett.\  {\bf 111}, 081601 (2013)
  doi:10.1103/PhysRevLett.111.081601
  [arXiv:1303.6292 [hep-lat]].

\bibitem{spinal} C.Miller and G.Krein, spin alignement in a pNJL model, private communication
\end{thebibliography}
\end{document}